\documentclass[aps,prb,a4paper,reprint,superscriptaddress,showpacs]{revtex4-1}
\usepackage{graphicx,graphics,color,epsfig}
\usepackage{epstopdf} %This line makes .eps figures into .pdf - please comment out if not required.
\usepackage{bm}
\usepackage{hyperref}
\usepackage{amsmath,subfigure}
\usepackage{amsfonts}
\usepackage{amssymb}
\usepackage{appendix}
\usepackage{booktabs}

\begin{document}

\title{Mechanism of Synergistic Effects of Neutron- and Gamma-Ray-Radiated PNP Bipolar Transistors}

\author{Yu Song}
\email{songyu@mtrc.ac.cn}
\affiliation{Microsystem and Terahertz Research Center, China Academy of Engineering Physics, Chengdu 610200, P.R. China}
\affiliation{Institute of Electronic Engineering, China Academy of Engineering Physics, Mianyang 621999, P.R. China}

\author{Ying Zhang}
\affiliation{Microsystem and Terahertz Research Center, China Academy of Engineering Physics, Chengdu 610200, P.R. China}
\affiliation{Institute of Electronic Engineering, China Academy of Engineering Physics, Mianyang 621999, P.R. China}

\author{Yang Liu}
\affiliation{Microsystem and Terahertz Research Center, China Academy of Engineering Physics, Chengdu 610200, P.R. China}
\affiliation{Institute of Electronic Engineering, China Academy of Engineering Physics, Mianyang 621999, P.R. China}

\author{Jie Zhao}
\affiliation{Microsystem and Terahertz Research Center, China Academy of Engineering Physics, Chengdu 610200, P.R. China}
\affiliation{Institute of Electronic Engineering, China Academy of Engineering Physics, Mianyang 621999, P.R. China}

\author{Dechao Meng}
\affiliation{Microsystem and Terahertz Research Center, China Academy of Engineering Physics, Chengdu 610200, P.R. China}
\affiliation{Institute of Electronic Engineering, China Academy of Engineering Physics, Mianyang 621999, P.R. China}

\author{Hang Zhou}
\affiliation{Microsystem and Terahertz Research Center, China Academy of Engineering Physics, Chengdu 610200, P.R. China}
\affiliation{Institute of Electronic Engineering, China Academy of Engineering Physics, Mianyang 621999, P.R. China}

\author{{Xiaofeng Wang}}
\affiliation{Microsystem and Terahertz Research Center, China Academy of Engineering Physics, Chengdu 610200, P.R. China}
\affiliation{Institute of Electronic Engineering, China Academy of Engineering Physics, Mianyang 621999, P.R. China}

\author{{Mu Lan}}
\affiliation{Microsystem and Terahertz Research Center, China Academy of Engineering Physics, Chengdu 610200, P.R. China}
\affiliation{Institute of Electronic Engineering, China Academy of Engineering Physics, Mianyang 621999, P.R. China}

\author{Su-Huai Wei}
\email{suhuaiwei@csrc.ac.cn}
\affiliation{Beijing Computational Science Research Center, Beijing 100193, China}

\begin{abstract}
The synergistic effects of neutron and gamma ray radiated PNP transistors are systematically investigated as functions of the neutron fluence, gamma ray dose, and dose rate. We find that the damages show a `tick'-like dependence on the gamma ray dose after the samples are radiated by neutrons. Two negative synergistic effects are derived, both of which have similar magnitudes as the ionization damage (ID) itself.
The first one depends linearly on the gamma ray dose, whose slope depends quadratically on the initial displacement damage (DD) and can be attributed to the healing of neutron-radiation-induced defects in silicon.
The second one has an exponential decay with the gamma ray dose, whose amplitude shows a rather strong enhanced low-dose-rate sensitivity (ELDRS) effect and can be  attributed to the passivation of {neutron-induced defects} near the silica/silicon interface by the gamma-ray-generated protons in silica, which can penetrate the silica/silicon interface to passivate the neutron-induced defects in silicon.
The simulated results based on the proposed model match the experimental data very well, but differ from previous model, which does not assume annihilation or passivation of the displacement defects.
The unraveled defect annealing mechanism is important because it implies that 
{displacement damages} can be repaired by gamma ray radiation or proton diffusion, which can have important device applications in the space or other extreme environments.
\end{abstract}

%\pacs{ }
\date{January 03, 2019} %\today
\maketitle

\section{Introduction}

Radiating particles 
such as gamma ray and neutrons %\textcolor{red}{in space and other extreme environments}
lead to ionizing (generating carriers) and
non-ionizing (generating atomic displacements) energy depositions in semiconducting materials, respectively.
As a result, the radiation damages of semiconductor devices contain both
{ionization damage (ID) and displacement damage (DD)}. In an environment with both gamma ray and neutron radiations,
it is often assumed that the total damage is a simple sum of ID and DD.
However, recent experiments have demonstrated that, the total damage in bipolar devices could be
{either} smaller or bigger than the simple sum of ID and DD, that is, there is
a negative or positive synergistic effect~\cite{Barnaby2001proton,Barnaby2002Analytical,gorelick2004effects,Li2011Combined,
Li2012Synergistic,Li2012Simultaneous,Li2015SynergisticEffect,
Li2015Research,Wang2015Simulation,Wang2016Ionizing,
li2016research,li2018synergistic}. 
However, the underlying mechanism of the synergistic effects is still not clear.
It has been speculated that the gamma-ray-induced charged traps in silica near the silica/silicon interface can change
the non-radiative Shockley-Read-Hall (SRH) carrier recombination by modifying charge distributions around the neutron-radiation-induced defects in the base region~\cite{Barnaby2001proton,Barnaby2002Analytical}. 
For example, in PNP bipolar transistors, the accumulation of the positive oxide trapped charge {($N_{ot}$)} near the interface increases
the electron density near the base surface by Coulomb attraction (see Fig. 1a).
As a result, the SRH recombination current is suppressed because of the widened carrier density difference in the base region, i.e., a negative synergistic effect arises. 
In NPN devices, a positive synergistic effect would arise because the positive oxide trapped charge lowers the hole density near the base surface
through Coulomb repulsion, which reduces the difference
of carrier densities in the base region, thus enhances the SRH recombination current. 
It has also been speculated that the gamma-ray-induced interface traps on the silica/silicon interface can also modify the charge distribution and change the SRH recombination.
However, since the interface traps are negative (positive) in N-type (P-type) base region, 
they always induce a positive synergistic effect for both PNP and NPN transistors~\cite{Li2012Synergistic,Li2012Simultaneous,Li2015SynergisticEffect}.
Very recently, it is deduced from deep level transient spectroscopy (DLTS) signals that the positive synergistic effect  in NPN transistors can be enhanced by the displacement defects generated in the oxide layer by low-energy proton radiations, because the defects can induce more oxide trapped charge during sequential high-energy proton radiations~\cite{li2016research,li2018synergistic}.  
All these models assume that in these processes, the  neutron-radiation-induced defect density in the base region is not affected by the gamma ray radiation.
However, no systematic study has been carried out to confirm the assumption and moreover, it is not clear what is the dependence of the synergistic effects as functions of the neutron fluence, gamma ray dose, and dose rate.

\begin{figure*}[!t]
  \centering
%\begin{subfigure}[b]{width=0.49\textwidth}
\centering
\includegraphics[width=0.49\textwidth]{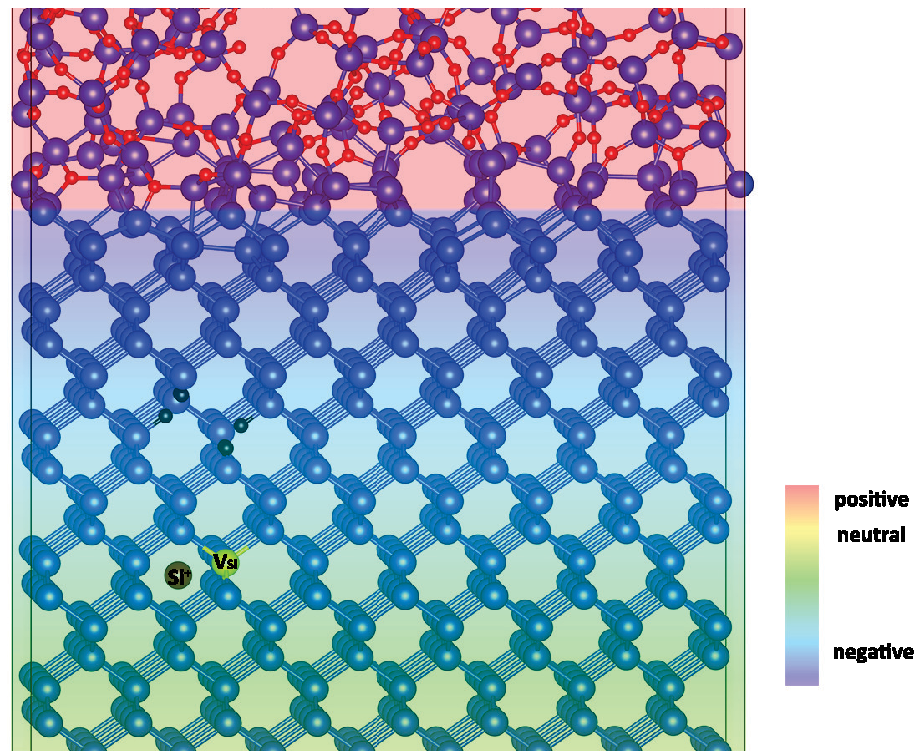}
%\caption{}
\label{}
%\end{subfigure}%
~
%\begin{subfigure}[b]{width=0.49\textwidth}
\centering
\includegraphics[width=0.49\textwidth]{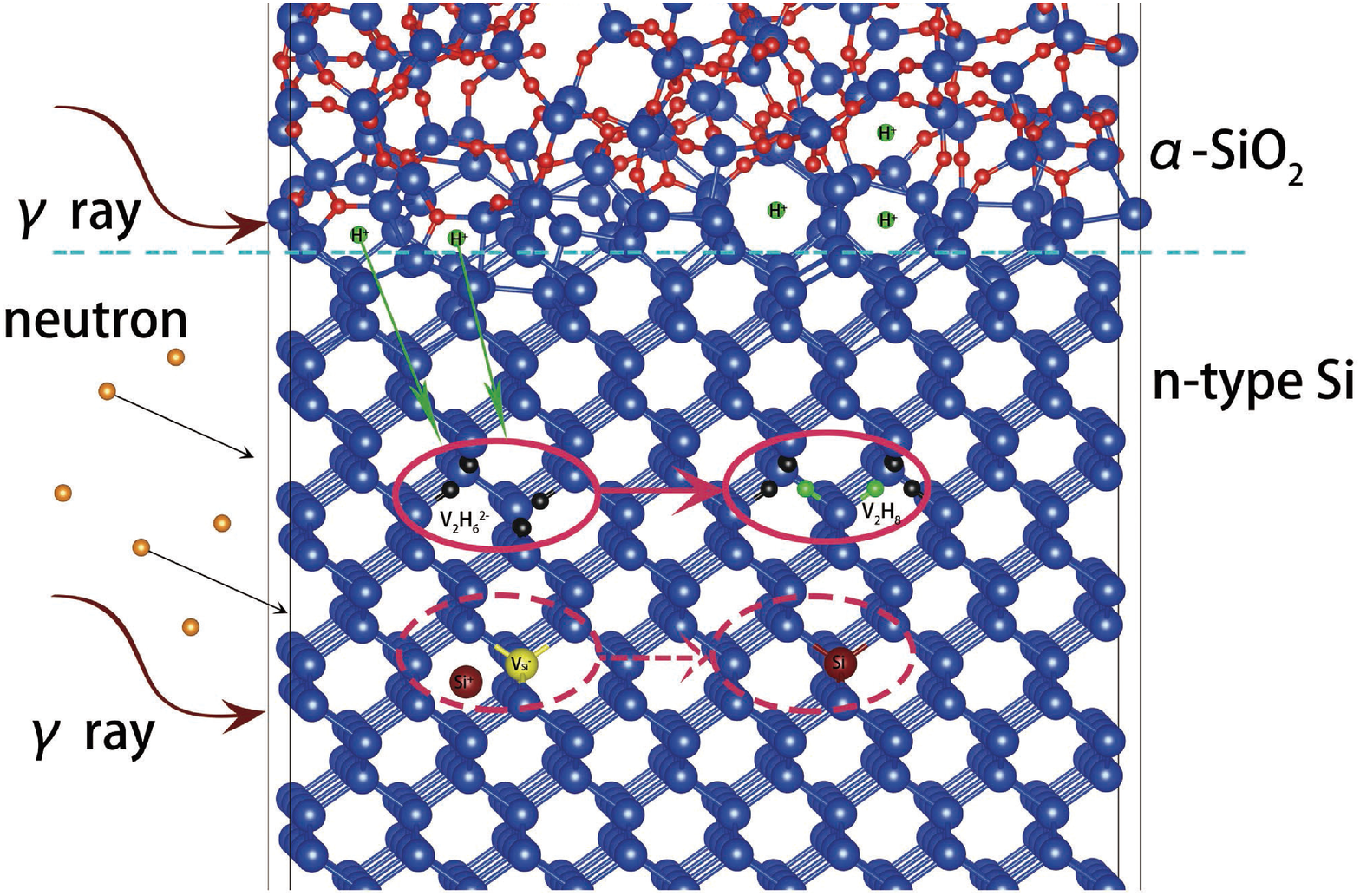}
%\caption{}
\label{}
%\end{subfigure}%
% \includegraphics[width=0.7\linewidth]{mechanism-4}
  \caption{
(color online) Schematic diagram for the mechanisms
of negative synergistic effect in PNP transistors.
(a) One of the traditional models~\cite{Barnaby2001proton,Barnaby2002Analytical} in the literature: the Coulomb interaction of gamma-ray-induced oxide trapping
charge in silica (red background) on the majority charge carriers (electrons) in silicon (green and blue background),
which leads to a change in the concentration of charge carriers and SRH recombination rate in silicon.
(b) The unraveled mechanism in this work: the gamma-ray radiation can heal the neutron-radiation-induced defects (vacancies and interstitials, dashed circle) and the induced protons in silica can also penetrate the interface to passivate the defects in silicon near the silica/silicon interface {($V_2H_6$ etc., solid circle)},
which lead to a decrease of the concentration of the SRH recombination centers in silicon.
  }\label{fig:decomposition}
\end{figure*}

In this work, we carry out systematic investigation on
the mechanism of the negative synergistic effects observed in PNP bipolar transistors.
The input-stage PNP transistors in the widely used operational amplifiers LM324N are successively radiated by
neutron and gamma ray with different neutron fluence, gamma ray dose, and dose rates.
We observe a `tick'-like dependence on the gamma ray dose on samples radiated by neutrons, from which we identified two negative synergistic effects that have magnitudes comparable with the ID itself.
The first one depends linearly on the gamma ray dose, whose slope depends quadratically on the initial DD.
The second one has an exponential decay with the gamma ray dose, whose amplitude shows a rather strong enhanced low-dose-rate sensitivity (ELDRS) effect.
To explain the observed data, we propose a \emph{defect annealing} model containing two terms,
a carrier-induced defect annihilation in silicon and a proton-induced defect passivation near the silica-silicon interface, which are schematically shown in Fig. 1(b).
The simulated results based on the proposed model match the experimental data very well, but differ from previous model that does not consider defect annihilation or passivation. Our unraveled mechanism is important because it implies that, we can repair damaged silicon devices used in space or under other extreme environments
by applying appropriate gamma radiation or proton diffusion.

The paper is organized as following.
In Sec. II, we describe the experimental setup of  the neutron-gamma radiations.
In the following Sec. III A, we first show
the data, which are found to display clear
`tick'-like damage-dose profiles.
We then demonstrate the presence of
two negative synergistic effects
in Sec. III B.
 {We then analyze the origin of the
two negative synergistic effects
in Sec. III C
and Sec. III D, respectively.}
The relative strength of the
two negative synergistic effects and
experimental conditions for the `tick'-like profiles
are investigated in the Sec. III E and Sec. III F, respectively.
The conclusion is made in Sec. IV.

\begin{figure}[!b]
\centering
\includegraphics[width=0.99\linewidth]{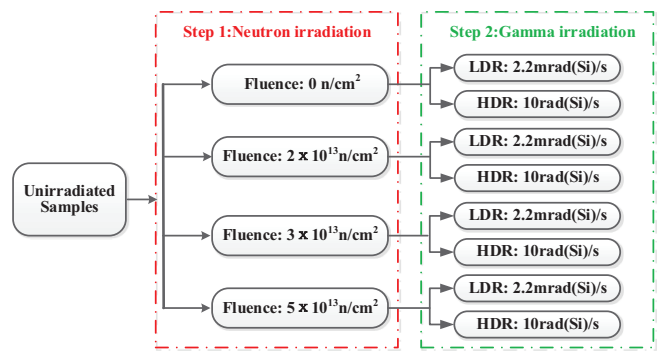}
\caption{(color online) %room-temperature annealing for 15 days?
Flowchart of the neutron/gamma radiation experiments.
For each of the eight radiation conditions, {2 chips with 8 PNP transistors} are used.
}\label{fig:expflow}
\end{figure}

\section{Experimental setup}\label{sec:model}

To investigate the behavior of the negative synergistic effect in PNP transistors,
LM324N chip (Texas Instruments, TI) with
4 operational amplifiers in each chip was selected for this study.
This is because the input stage of the operational amplifier is very straightforward hence the input bias current is directly related to the base current of the input-stage PNP transistors~\cite{pease1996compendium,barnaby1999identification}.
Based on this fact, 
the synergistic effect of the input bias current of an LM124 chip, which is very similar to LM324N, has been understood and modeled in the level of the input-stage transistors~\cite{Barnaby2001proton,Barnaby2002Analytical}.
On the other hand, %we use LM324N also because 
with a substrate strucuture and lightly doping, the input-stage
transistors in LM324N are sensitive to both neutron and gamma ray radiations,~\cite{Barnaby2001proton,Barnaby2002Analytical}
which is essential for observing a remarkable synergistic effect.
The processes of the experiments are shown in Fig.~\ref{fig:expflow}.
Six neutron-gamma conditions are employed:
first neutron radiation
with the fluence of 2$\times$10$^{13}$/cm$^2$, 3$\times$10$^{13}$/cm$^2$, and
5$\times$10$^{13}$/cm$^2$, respectively;
then gamma radiation to 5krad(Si) at a low dose rate of 2.2 mrad(Si)/s
and a high dose rate of 10 rad(Si)/s, respectively.
Another two pure gamma radiation conditions are used to obtain
the artificial damage.
For each condition, we use  {2 chips (i.e., 8 PNP transistors)}.
In all the experiments, chips were radiated in an unbiased configuration with all pins shorted.
The input bias currents
were measured by BC3193 discrete semiconductor testing systems
and used to analyze the damages of the input-stage PNP transistors.
Neutron radiations were performed at the Chinese Fast Burst Reactor-II (CFBR-II) of
Institute of Nuclear Physics and Chemistry, China Academy of Engineering
Physics, which provides a controlled 1MeV equivalent neutron radiation.
Gamma ray radiations were done at
College of Chemistry and Molecular Engineering of Peking University.

\section{Results and discussion}

\subsection{`Tick'-like dependence on the gamma ray dose}

\begin{figure}[!t]
\centering
\includegraphics[width=0.95\linewidth]{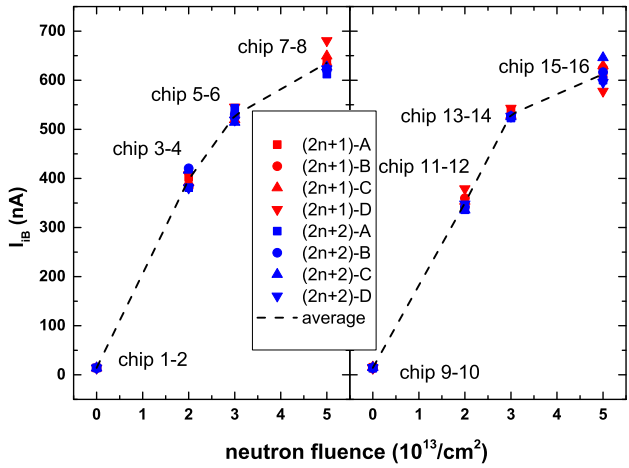}
\caption{(color online) %the laber is wrong.
The input bias current of the input-stage PNP transistors in
LM324N %in unit of nA
as a function of the neutron fluence. %in unit of n cm$^{-2}$.
{The measurements are performed 15days after the neutron radiations are over.
The temperature during radiation, storage, and measurements are 20$^\circ$C.}
For each fluence (n=0-7), {the two chips are distinguished by different colors
of red and blue, respectively;}
in each chip, the four transistors (labeled as A, B, C, and D) are distinguished by different symbol shapes of square, circle, up-triangle, and down-triangle, respectively.
}\label{fig:DDradiation}
\end{figure}

The samples are first radiated by neutrons.
Fig. 3 shows the pure DD response of the studied devices.
The transistor numbers are indicated in the figure.
Also plotted is the average value of {the 8 PNP transistors} for each neutron fluence.
It is seen that, the input bias current ($I_{iB}$) increases sub-linearly
with the neutron fluence.
When the neutron fluence accumulates from 0 to 5$\times$10$^{13}$/cm$^2$,
the average bias current increases from about ~10 nA to about ~600 nA.
Due to the different sample quality, the DD of 8 transistors under a same fluence are also {different by tens of nA},
which will lead to different latter gamma ray response, as discussed below.
The thin oxide layer in the PNP transistor is almost transparent for neutrons~\cite{raymond1987comparison}.
Neutron radiation has been shown to introduce acceptor-like defects in
Si~\cite{Li1995Experimental,Lutz1995Simplistic,Schulz1994Long,
Li1994Modeling}.
The divacancies ($V_2$) and vacancy-oxygen (VO) pairs prominently identified in DLTS measurements~\cite{watts1996new,Lindstrom2001Radiation}
are thought to be the candidates for these negative charged defect centers~\cite{Li1994Modeling,Myers2008Model,watts1996new}, see Fig. 1(b).
Accompanying, Si self-interstitials are also generated.
The lifetime of minority carrier ($\tau$) is inversely proportional to the concentration of defects.~\cite{pierret1987advanced}
Accordingly, the generation of defects with an increase of neutron fluence results in a persistent decrease of the lifetime
of minority carrier and hence the DD degradation,~\cite{adell2014dose}
\begin{equation}
\Delta I_{iB}^{D} = \frac{q n_i A x_{dB}}{2 \tau} e^{\frac{q V_{BE}}{2 k_B T}}~.
\end{equation}
Here the superscript $D$ stands for DD, $q$ is the charge of minority carrier, 
$n_i$ is the concentration of intrinsic carriers,
$A$ and $x_{dB}$ are the area and depth of the space charge region, respectively, 
$V_{BE}$ is the bias between the base and emitter electrodes, 
and $T$ is the temperature.
This relation implies that the concentration of the generated defects is proportinal to the increase of the input bias current, $[V] = \lambda \Delta I_{iB}^{D}$, where $ \lambda^{-1} \propto \frac{q n_i A x_{dB}}{2} e^{\frac{q V_{BE}}{2 k_B T}}$.
From the data as shown in Fig.~3, we can see that 
the concentrations of the generated defects in silicon are different
for different neutron fluence and samples.

\begin{figure*}[!t]
%\centering
%\subfigure{\label{(a)}}%[b]{0.22\textwidth}
\centering
\includegraphics[width=0.22\textwidth]{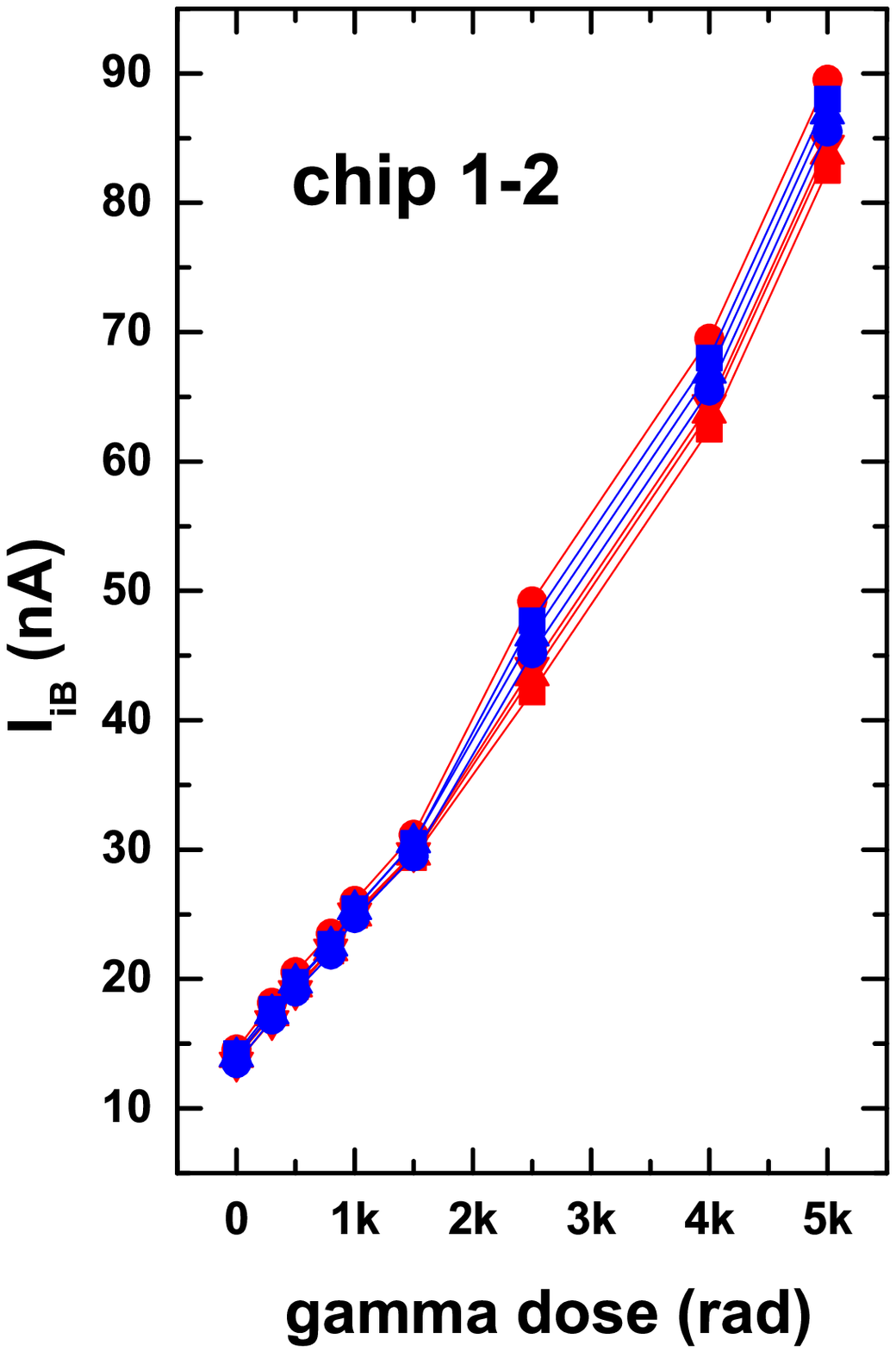}%\\
%\caption{}
%\end{subfigure}%
%~
%%\begin{subfigure}[b]{0.23\textwidth}
%\centering
\includegraphics[width=0.23\textwidth]{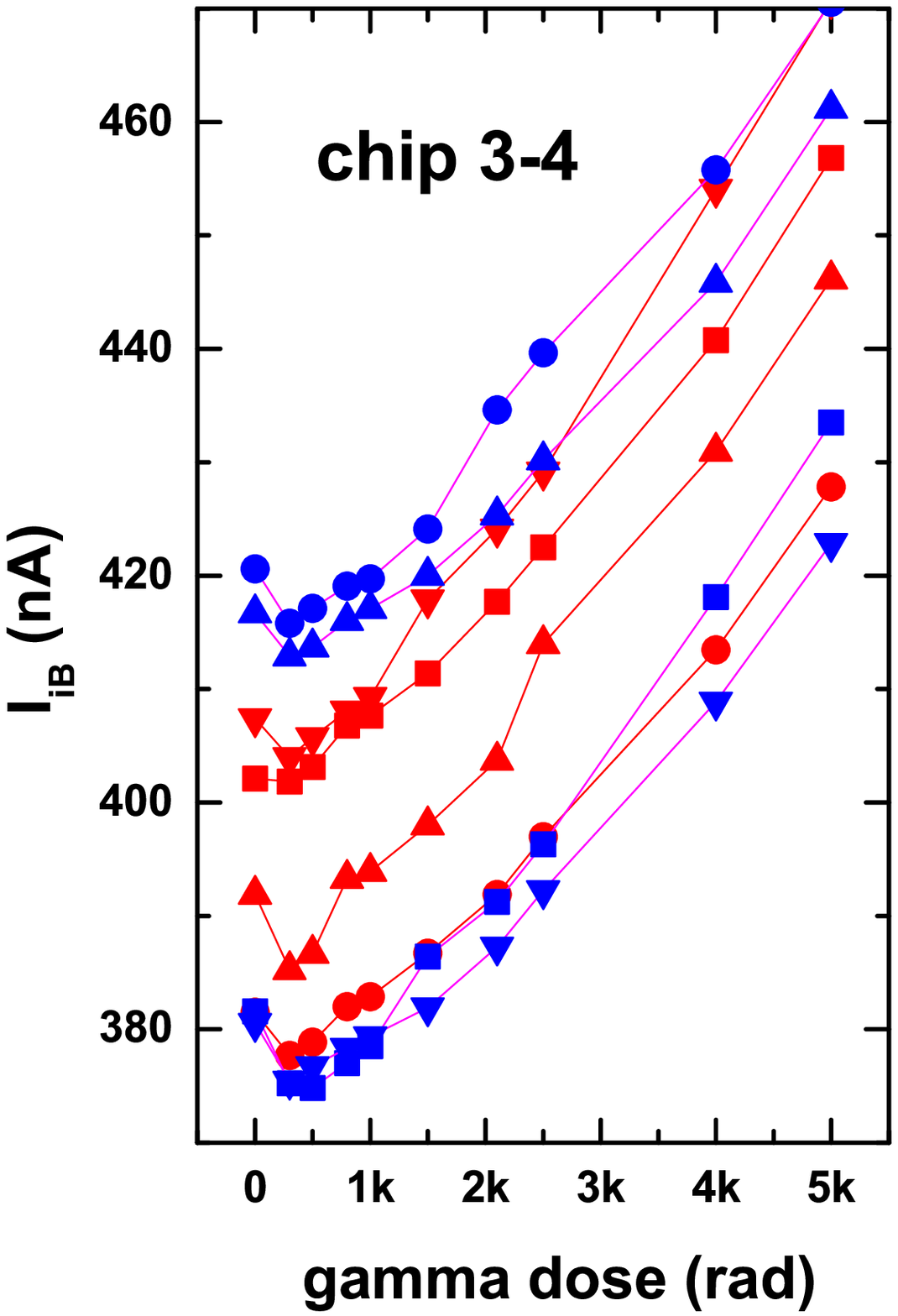}%\\
%\caption{}
%\label{}
%%\end{subfigure}%
%~
%%\begin{subfigure}[b]{0.23\textwidth}
%\centering
\includegraphics[width=0.23\textwidth]{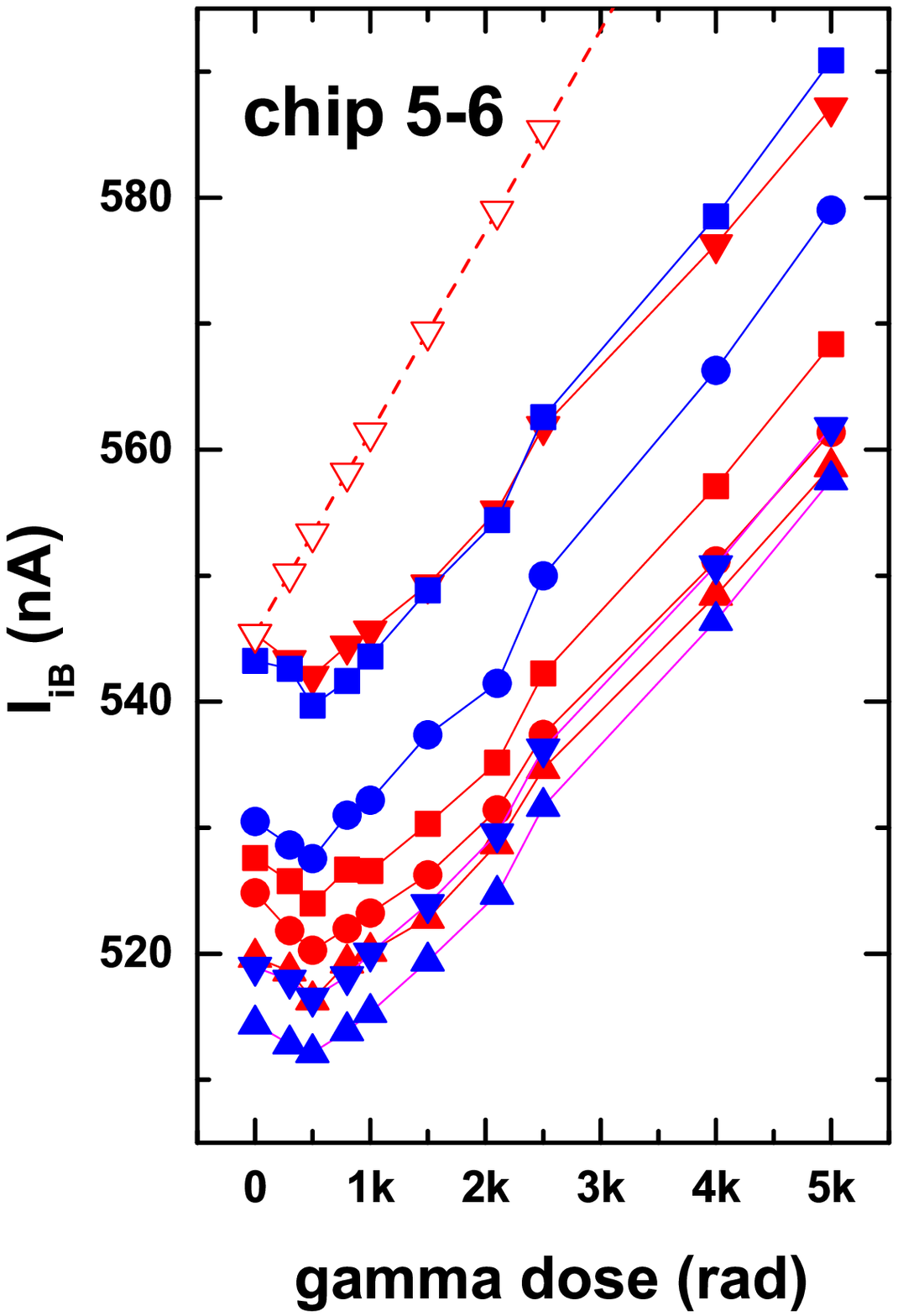}%\\
%\caption{}
%\label{}
%%\end{subfigure}%
%~
%\begin{subfigure}[b]{0.23\textwidth}
%\centering
\includegraphics[width=0.23\textwidth]{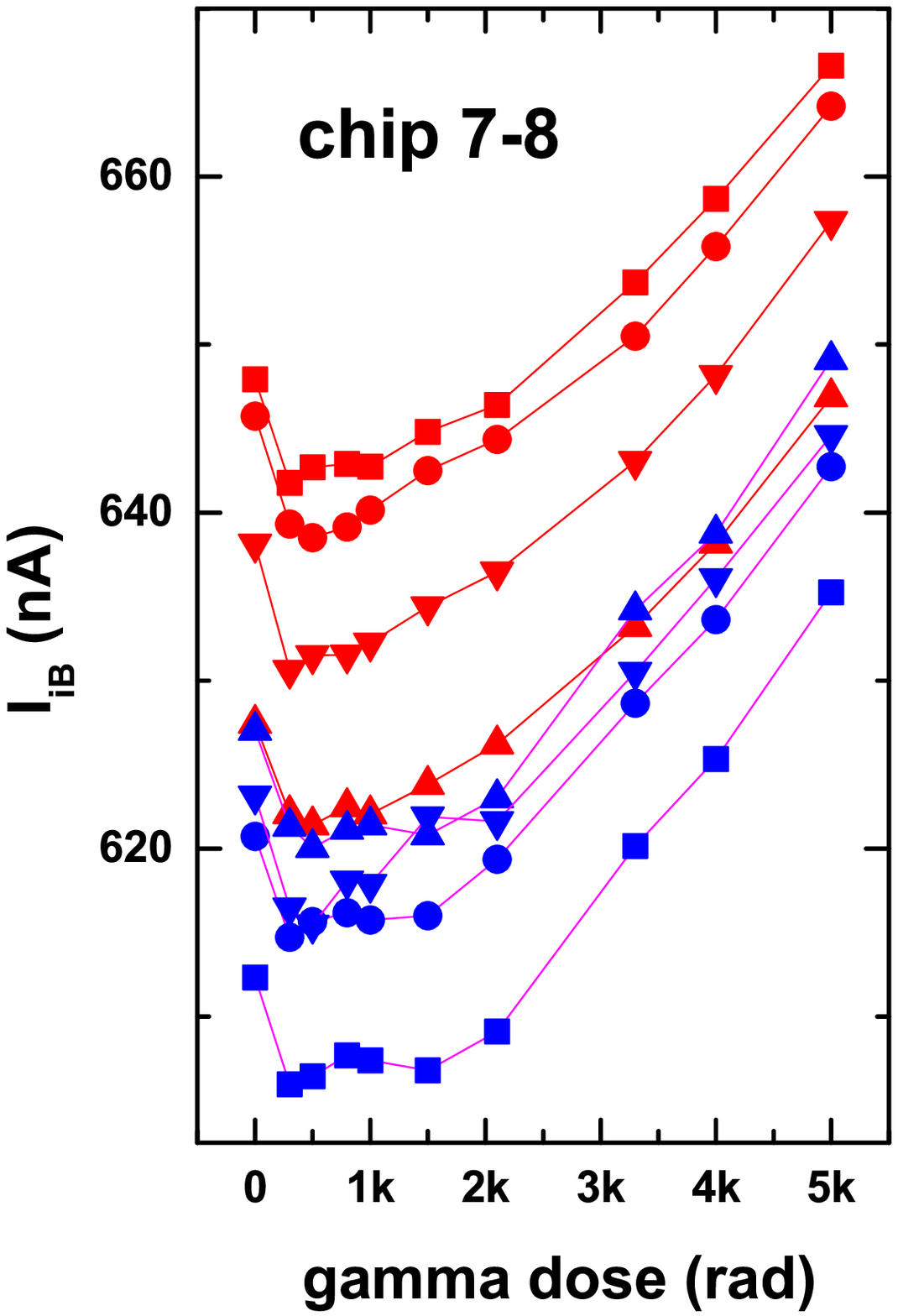}
%\caption{}
%\label{}
%\end{subfigure}%
%~
\\
%\begin{subfigure}[b]{0.22\textwidth}
%\centering
\includegraphics[width=0.22\textwidth]{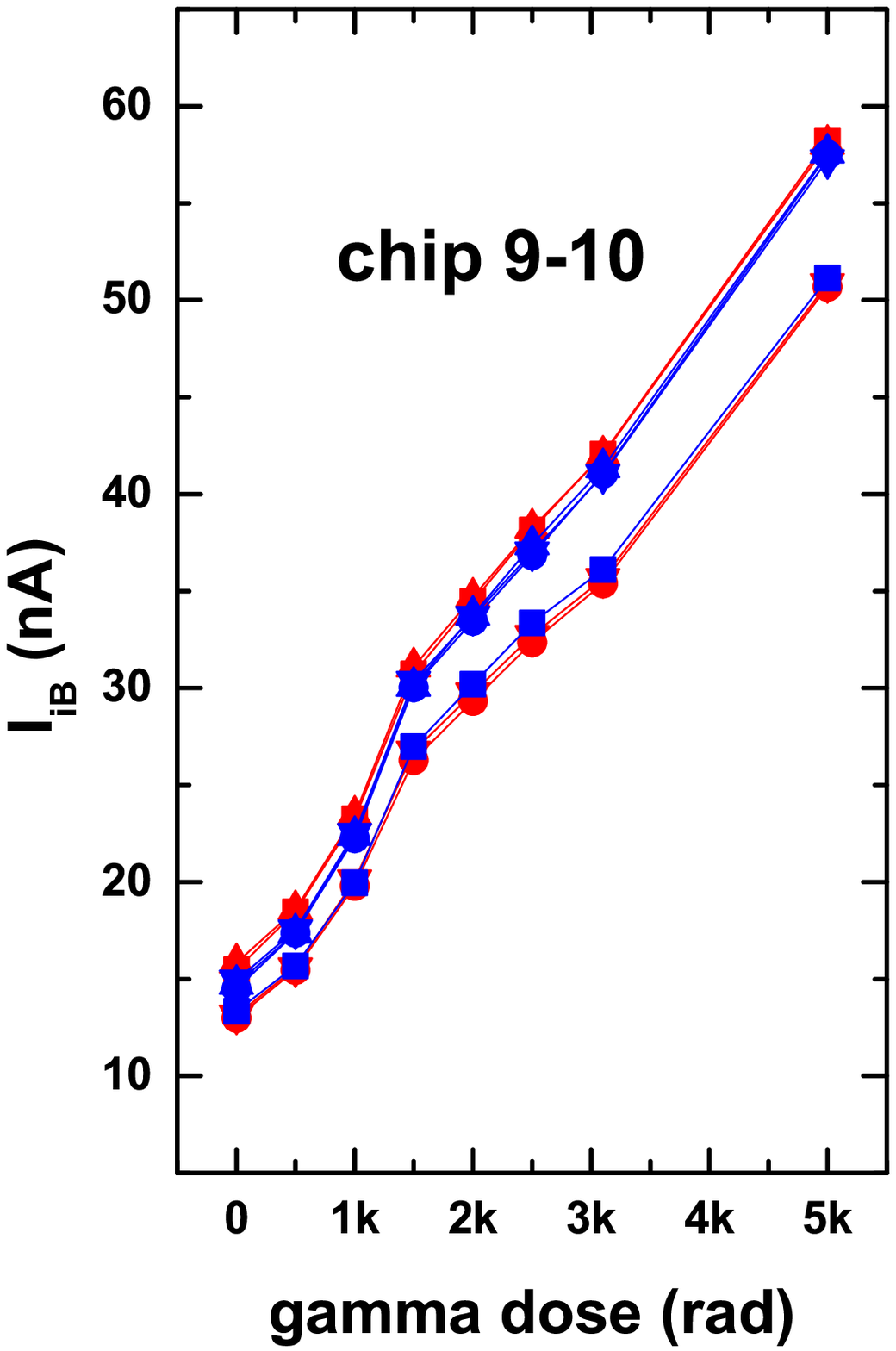}%\\
%\caption{}
%\label{}
%\end{subfigure}%
%~
%\begin{subfigure}[b]{0.23\textwidth}
%\centering
\includegraphics[width=0.23\textwidth]{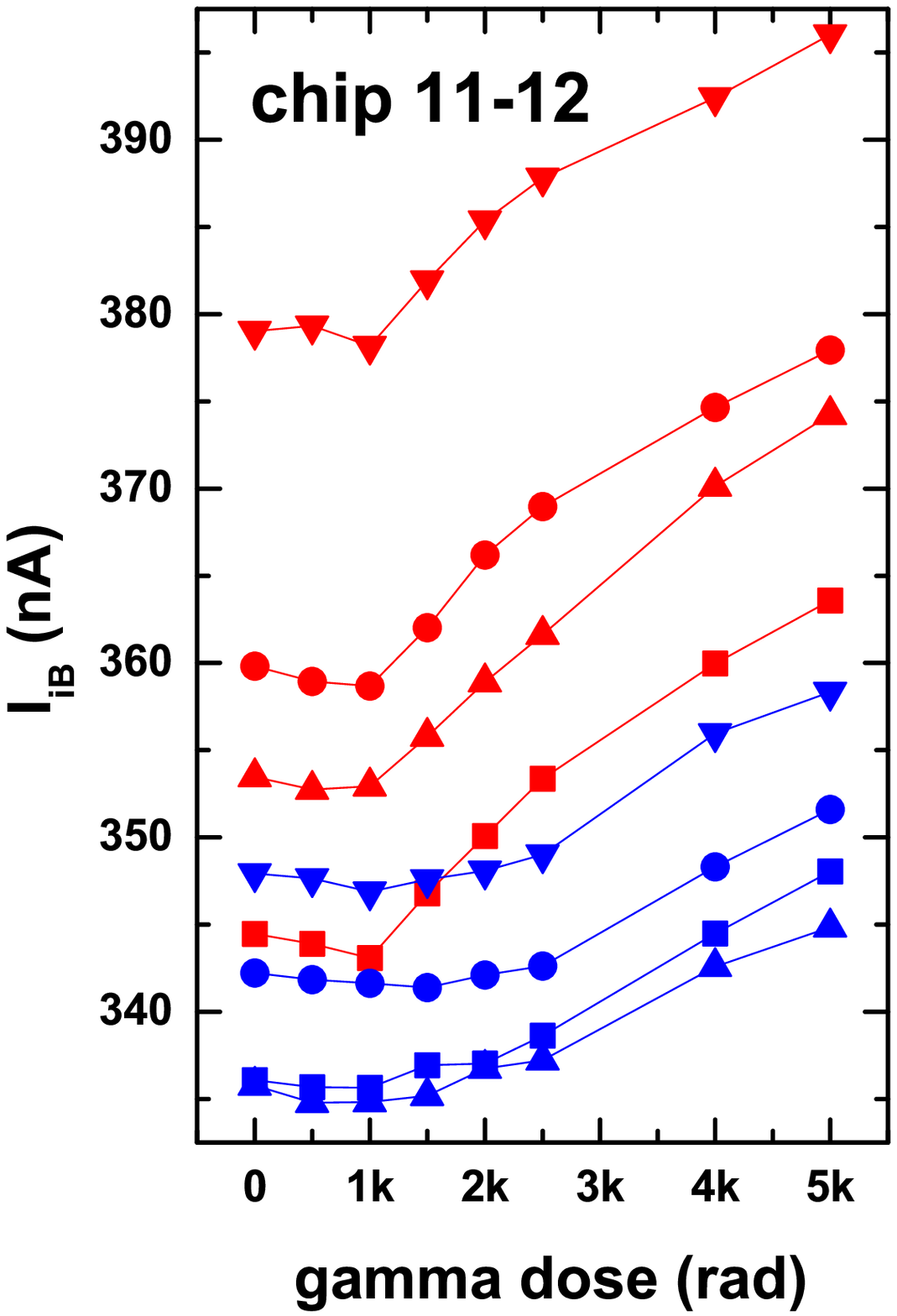}%\\
%\caption{}
%\label{}
%\end{subfigure}%
%~
%\begin{subfigure}[b]{0.23\textwidth}
%\centering
\includegraphics[width=0.23\textwidth]{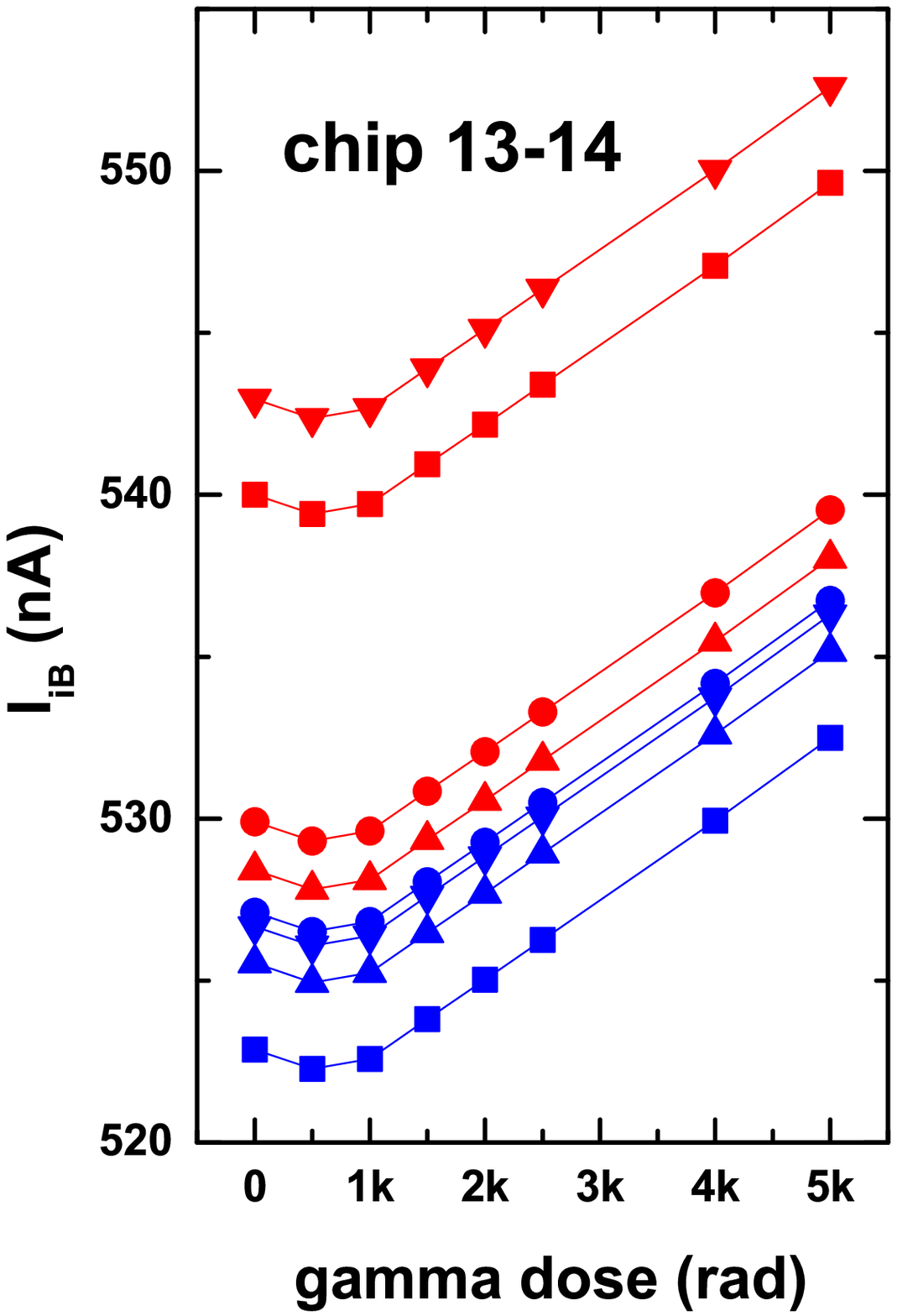}%\\
%\caption{}
%\label{}
%\end{subfigure}%
%~
%\begin{subfigure}[b]{0.23\textwidth}
%\centering
\includegraphics[width=0.23\textwidth]{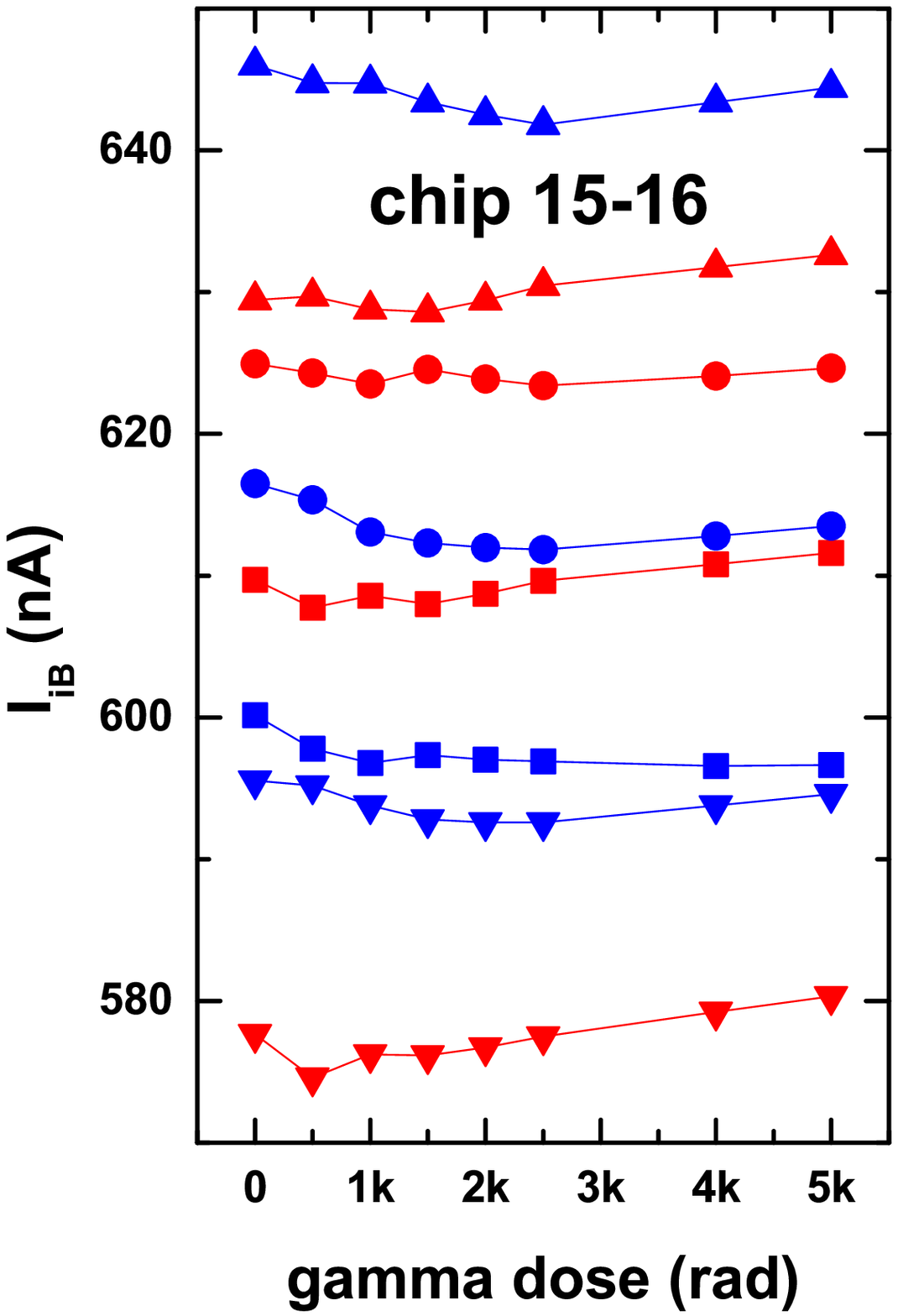}
%\caption{}
%\label{}
%\end{subfigure}%
%~
\caption{(color online)
(a)-(d) The input bias current as a function of the gamma ray dose
at low dose rate of  $2.2~\text{mrad(Si)/s}$
for PNP transistors under radiations of different initial neutron fluence.
From left to right, the neutron fluence is
$0$, $2\times 10^{13}~\text{cm}^{-2}$, $3\times 10^{13}~\text{cm}^{-2}$, and $5\times 10^{13}~\text{cm}^{-2}$, respectively.
(e)-(h) The same configurations as (a)-(d) but at a high dose rate of $10~\text{rad(Si)/s}$.
In each split, {the two chips are distinguished by different colors
of red and blue, respectively;}
in each chip, the four transistors are distinguished by different symbol shapes of square, circle, up-triangle, and down-triangle, respectively.
%The `bad' samples are distinguished by dashed curves.
} \label{fig:TIDafterDDradiation}
\end{figure*}

To obtain the simple sum of DD and ID, {chips No. 1-2 and No. 9-10}
are radiated by gamma ray with low and high dose rate, respectively.
The results are displayed in Fig. 4(a) and 4(e), respectively.
It is seen that,
the ID increases almost linearly with the {gamma ray dose},
i.e., $\Delta I_B^{I} = k_0 x$, {where $x$ is the gamma ray dose in unit of krad(Si).}
An average coefficient of $k_0^L = 16.0$ nA/krad(Si)
and $k_0^H = 7.0$ nA/krad(Si)  are obtained
for the low and high dose rate, respectively.
There is a clear {ELDRS effect}~\cite{pease2008eldrs}, with an enhancement factor of about 2.3.
%It is well-known that, 
Under gamma ray radiations,
protons are generated in the silica layer through
dissociation reactions between oxide trapping
charges and hydrogen molecule~\cite{rowsey2011quantitative,yue2018dissociation},
see Fig. 1 (b).
The generated protons diffuse to the oxide/silicon interface and further generate interface traps ($N_{it}$) through depassivation reactions on passivated defects Si-H bonds on the interface~\cite{rashkeev2001defect,rowsey2011quantitative}.
The interface traps result in an increase in the surface recombination velocity ($\Delta s$)  above the base region, which leads to the base current {increment}
~\cite{schmidt1996modeling,kosier1995physically} % tolleson2018improved
\begin{equation}
\Delta I_{iB}^{I} = \Delta s \frac{q n_i P_E x_{dB}}{2} e^{\frac{q V_{BE}}{2 k_B T}}~.
\end{equation}
Here  the superscript $I$ stands for ID, $\Delta s = v_{th} \sigma \Delta N_{it}$,
where $v_{th}$ is carrier thermal velocity and $\sigma$ is the carrier capture cross section.
$P_E$ is the emitter perimeter.
From the data in Fig. 4 (a) and (e), it is seen that $N_{it}$ increases linearly with the gamma ray dose.
The concentration of the generated protons is almost constant, which is
higher for the low dose rate case.

The simple sum of the DD and ID damages then reads:
\begin{equation}
\Delta I_{iB}^{D+I} = D_0 +k_0^{L(H)} x,
\end{equation}
where $D_0$ is the initial DD.
The result is plotted in Fig. 4(c) for {transistor No. 5D}.
%%%%
The low-dose-rate gamma response of {chips No. 3-8 (24 transistors)}
with various $D_0$  are
shown in Fig.~\ref{fig:TIDafterDDradiation} (b-d).
Similarly, the high-dose-rate gamma response of {chip No. 11-16
(another 24 transistors)} are shown in  Fig.~\ref{fig:TIDafterDDradiation} (f-h).
It is seen that, {for all 48 samples} the total damages are smaller than the {simply} summed ones.
In other words, a clear negative synergistic effect is observed.
In previous works~\cite{Barnaby2001proton,Barnaby2002Analytical}, this effect is attributed to the change of the electron density
in silicon caused by positive oxide charge accumulation in silica, see Fig. 1(a). 
However, in this work, our analysis of the features in the data suggest
that these effects are due to the annihilation and passivation of the DD defects in silicon.

%%%%%%%%%%%%%%%%%%%%%%%%%%%%%%%%%%%%%%%%%%%%
\begin{figure}[!t]
\centering
% %\begin{subfigure}[b]{0.24\textwidth}
% \centering
\includegraphics[width=0.95\linewidth]{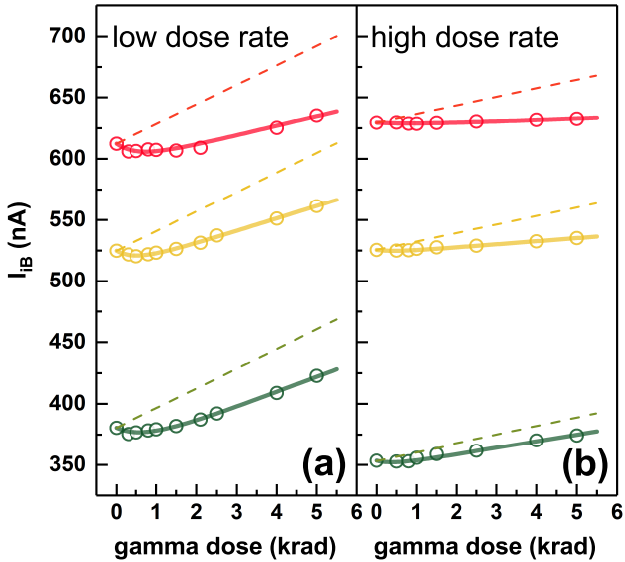}
\caption{%add the artificial line!
(color online) The experimental data (dots) and fitting curves (solid) of the input bias current as function of gamma ray dose of samples with various initial DD. %subtract the component of IDD.
(a) The case of transistors radiated with low dose rate. (b) The case of transistors radiated with high dose rate.
The simple combined damages are plotted by the dashed lines.
} \label{fig:TIDpure}
\end{figure}
%%%%%%%%%%%%%%%%%%%%%%%%%%%%%%%%%%%%%%%%%%%%

The damages display a `tick'-like profile:
they abnormally decreases for small gamma ray dose %near-exponentially
and then increase almost linearly for large gamma ray dose.
However, the slope for the latter is obviously smaller than $k_0^L$
or $k_0^H$.
In the following, we will show that, they result from the \emph{proton passivation and carrier-induced annihilation
of the neutron-radiation-induced defects} in silicon, respectively.

\subsection{Linear and exponential negative synergistic effects and their fluence and dose rate dependence
}

To obtain a general trend as a function of the initial displacement damage and gamma ray dose rate,
in Fig. 5, we plot six typical damage-dose curves:
three with $D_0$ of about 380nA, 520nA, and 610nA
radiated with the low dose rate,
and three with $D_0$ of about 350nA, 520nA, and 630nA
radiated with the high dose rate.
It is seen that, the larger the initial DD, %or? the lower the dose rate,
the bigger the decrease in the slope at large gamma ray dose;
on the other hand, the lower the dose rate, the stronger the abnormal decrease of
the damage at small gamma ray dose.

To further investigate the possible origin of these two synergistic effects, we
fit all the data for the low/high dose rate case in Fig. 4(b-d)/(f-h).
Considering both the linear behavior at the large gamma ray dose %in the damage-dose curves
and %decreasing velocity
the exponential-like decline at the small gamma ray dose in the damage-dose curves,
we find that the negative synergistic effects of all 48 curves with different 
nonzero $D_0$ in Fig. 4 can fit very well to the relatively simple function containing a negative linear term
and a negative exponential term,
\begin{equation}\label{eq:fitting}
\Delta I_{iB} = \Delta I_{iB}^{D+I} - a x  - b (1 - e^{- c x}).
\end{equation}
Here $a$ is the deviation slope of the linear term; it has a dimension of nA/krad.
$b$ has a dimension of nA and stands for the amplitude of the exponential decay term.
$c$ describes the decay rate; %as a function of dose.
its dimension is 1/krad.
The fitted curves for the six typical samples are shown in Fig. 5.
The individual linear synergistic term $-a x$ and
the exponential synergistic term $- b (1 - e^{- c x})$
for the 3 {curves with different $D_0$} in Fig. 5(a)
are shown in Fig.~6 for further analysis.
It is clear that, there are two separate negative synergistic effects in the investigated system.

\begin{figure}[!t]
\centering
\includegraphics[width=0.95\linewidth]{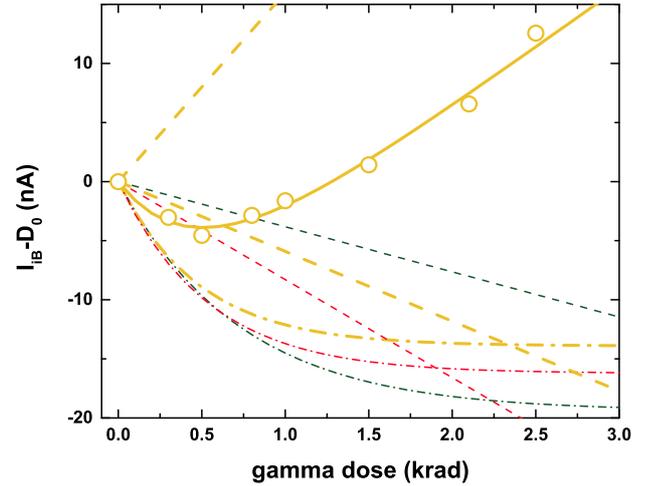}
\caption{
(color online) The decomposition of the damage-dose curve with medium DD (orange)
in {Fig. 5(a)}.
Dashed line is for the linear annealing effect
and dash-dot curve is for the exponential annealing effect.
For comparison, the linear and exponential terms for damage curves with low and high DD 
(green and red, respectively) in {Fig. 5(a)} are also plotted.
For clearness, the initial DD is deducted from the input bias currents.
}\label{fig:decomposition}
\end{figure}

%%%%%%%%%%%%%%%%%%%%%%%%%%%%%%%%%%%%%%%%%%%%
\begin{figure*}[!t]
\centering
\centering
\includegraphics[width=0.46\textwidth]{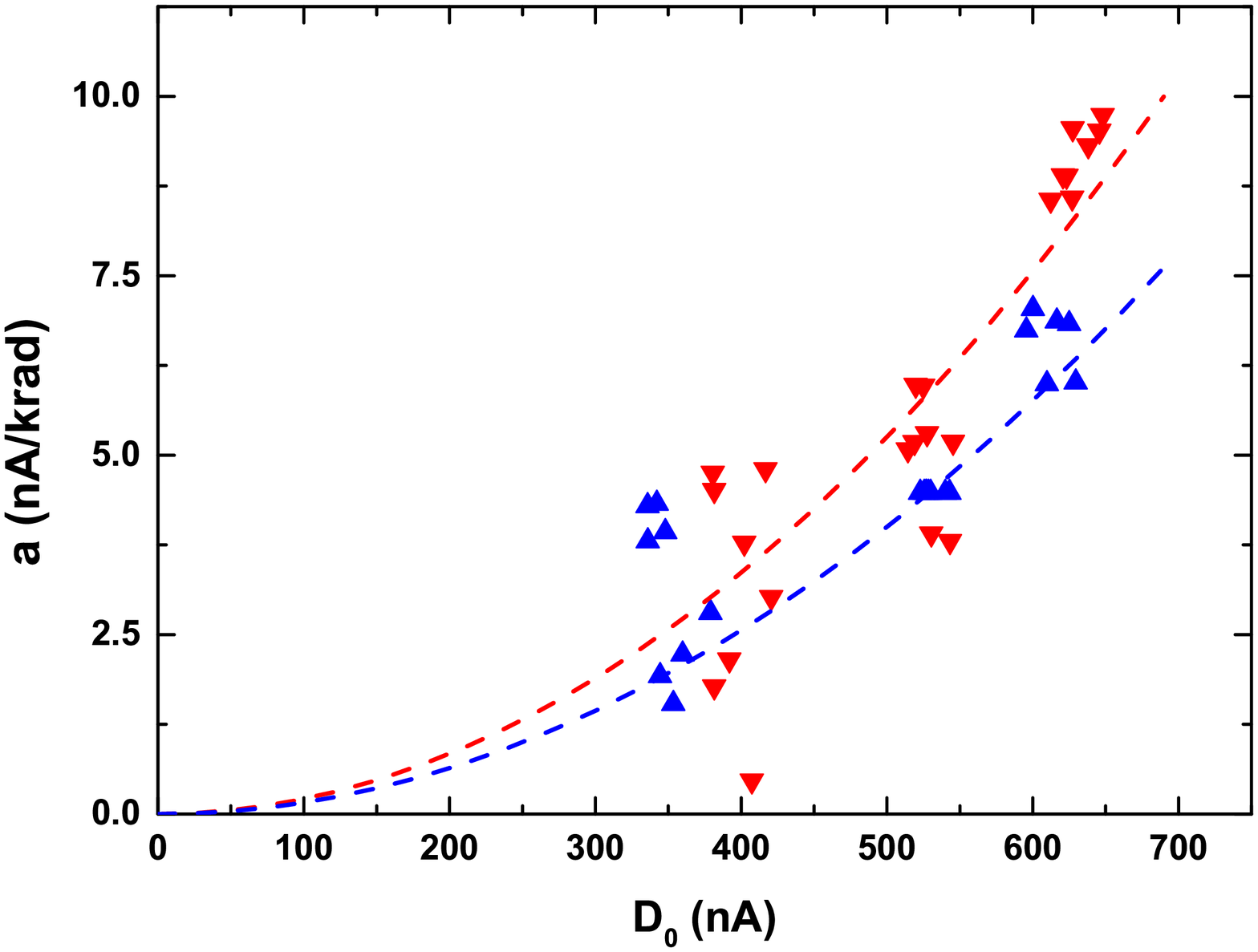}
%\caption{}
\label{}
%\end{subfigure}%
~
%\begin{subfigure}[b]{0.45\textwidth}
\centering
\includegraphics[width=0.45\textwidth]{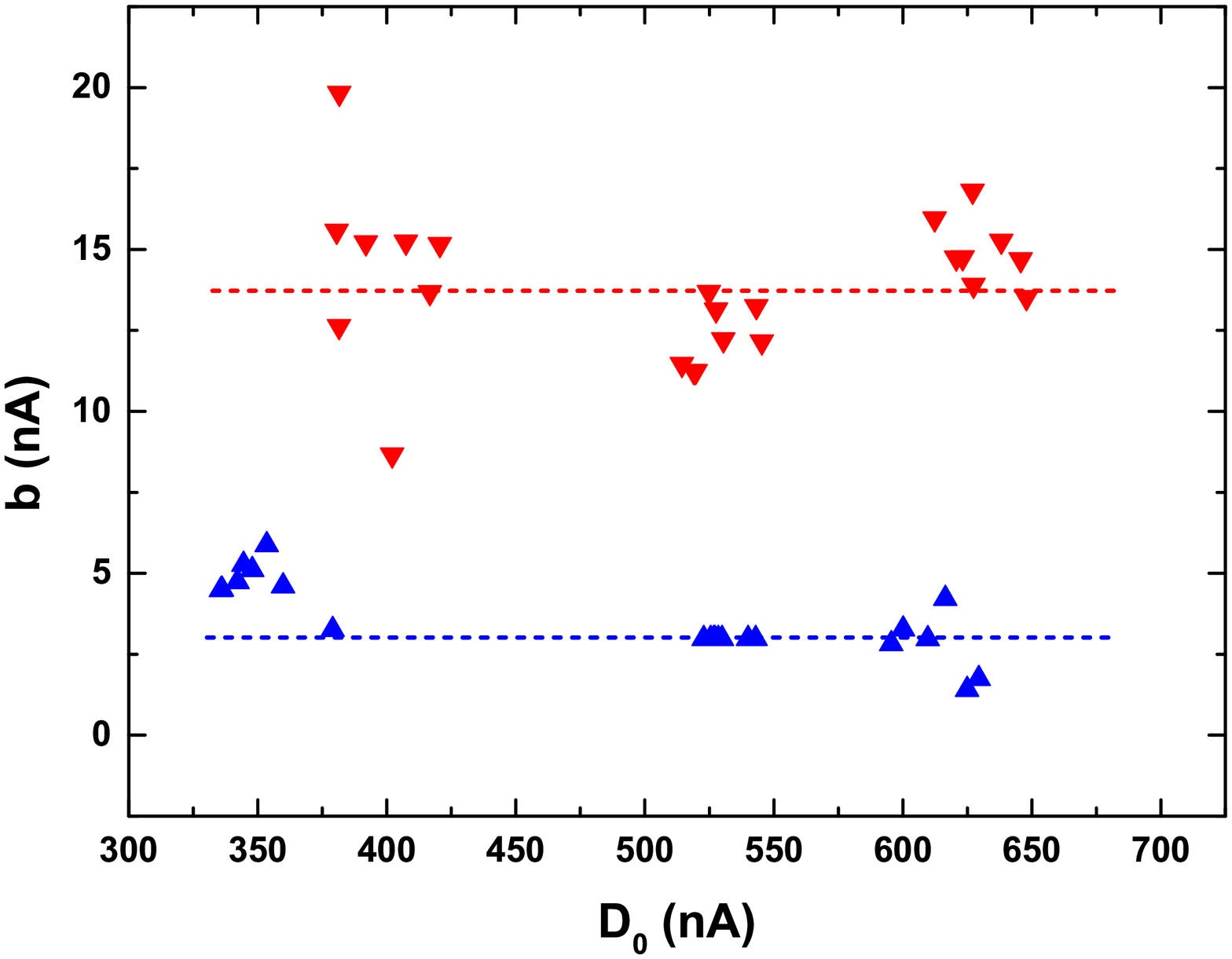}
%\caption{}
\label{}
%\end{subfigure}%
~
\caption{
(color online) (a, b) {The fitting parameters $a$ and $b$ in Eq.~(\ref{eq:fitting}) for the 48 curves in Fig. 4 as a function of the initial displacement damage.
The fitting value of $c$ is 2.1/krad for all samples.
Red for the low dose rate and blue for the high dose rate.
%(d, e) The re-fitting parameters $a$ and $b$ using a standard $c=2.1$/krad.
}
}\label{fig:linearterm_annealing}
\end{figure*}
%%%%%%%%%%%%%%%%%%%%%%%%%%%%%%%%%%%%%%%%%%%%

{The fitting decay rate ($c$) is found to be insensitive to the neutron
fluence or gamma ray dose rate. It is almost the same ($c=2.1$/krad) for all 48 damage-dose curves.}
The {fitting parameters $a$ and $b$} as a function of $D_0$ for
samples radiated at the low/high dose rate are shown {in Fig. 7 (a) and (b)}, respectively.
It is seen that, the decrease in the slope of the damage ($a$) increases with $D_0$ superlinearly.
Further fitting of the $a$-$D_0$ data shows that,
an interesting and simple relation is satisfied
for both the low and high dose rates,
\begin{equation}
a= \alpha D_0^2,
\end{equation}
where {$\alpha$=21/(mA$\times$ krad)  and $\alpha$=16/(mA$\times$ krad)} for the low and high dose rate case, respectively.
On the other hand, the amplitude of the exponential term ($b$) shows a strong dose rate dependence and is not very sensitive to the initial displacement damage.

\subsection{Dependence of the linear synergistic effect on the neutron fluence: carrier-induced defect annihilation in silicon
}\label{chap:linearannealing}

We first consider the origin of the linear synergistic effect.
According to Eqs. (4) and (5), the linear synergistic term can be re-written as
\begin{equation}\label{eq:linearannealing}
\Delta I_{iB}^{SE1} =  - \alpha D_0^2 x.
\end{equation}
This expression suggest the following.
1) The term depends exactly on $D_0$, which means that this
synergistic effect is related to the neutron-induced defects in silicon;
2) The term is negative, which means that the neutron-induced defects is reduced. In other words, \emph{an annealing effect of the neutron-induced defects in silicon} happens.
3) The term is a quadratic function of the defect concentration, which suggest that the annihilation of two {kinds} of related defects happens;
4) The term also depends linearly on the gamma ray dose, which means that the annihilation of the neutron-induced defects is induced by gamma-induced charge
carriers.

This observation clearly cannot be attributed to the charge redistribution in the silicon region as implied in early models.~\cite{Barnaby2001proton,Barnaby2002Analytical,
Li2012Synergistic,Li2012Simultaneous,Li2015SynergisticEffect}
Here, we propose that the defect annihilation can be described by the following annihilation reaction
\begin{equation}
V_n^- + I^+ =V_{n-1}~,
\end{equation}
{where $V$ is the vacancy and $I$ is the Si interstitial,
both of which are charged by the gamma radiation, {see Fig. 1(b)}. 
We note that the annihilation of defects in silicon due to  injected charge carriers has been found in the past
\cite{gregory1967injection,barnes1969thermal,harrity1970short} and the origin has been attributed to the enhanced mobility of defects through alternating capture and lose of electrons.~\cite{kimerling1975role,kimerling1976new,bar1984electronic,Bar-Yam1984,car1984microscopic}
Among various defects, the isolated Si interstitial defects are mobile particles}.
From this reaction, we can see that,
for a fixed gamma ray dose rate, the more the neutron-induced defects exist in the sample
the easier to find $I^+$ as well as $V_n^-$.
On the other hand, for a fixed defect concentration,
the more the charge carriers ([h]) are excited by the $\gamma$-ray,
the more the mobile defects are stimulated.
According to the chemical reaction rate equation, the annihilation rate of the vacancies reads 
$d[V_n^-]/dt = - k_a [V] [I] [h]$, 
where $k_a$ is the reaction rate constant between the excited vacancies and
interstitials.
A simple calculation gives %Integrate considering V0, 
$\Delta V_n^- = - k_a [V] [I] x$, where $x= [h] t$ is the gamma ray dose.
Here the defect concentrations on the right side in the reaction rate equation are much larger than the change and
are regarded as constants during the gamma ray radiation. 
Recalling the relation between the defect concentration and DD,
$[V] = [I] = \lambda D_0$, the above relation can be rewriten as
$\Delta I_{iB}^{SE1} = -k_a  \lambda D_0^2 x$.
This is just Eq. (6), with $k_a \lambda$ corresponding to $\alpha$.
As mentioned above, the value is a slightly larger for the low dose rate case
(21/(mA$\times$ krad)) than for the high dose rate
(16/(mA$\times$ krad)),
with an enhancement factor of about 1.3.
This means that there is also an ELDRS effect in the linear
synergistic effect.
It can be conclued that, the square law come from the fact that, the annihilation requires the existence of both interstitial and vacancy;
while the interstitial defect caused by the neutral radiation has the same quantity as the vacancy defect.

\begin{figure*}[!t]
\centering
%\begin{subfigure}[b]{0.45\textwidth}
\centering
\includegraphics[width=0.45\textwidth]{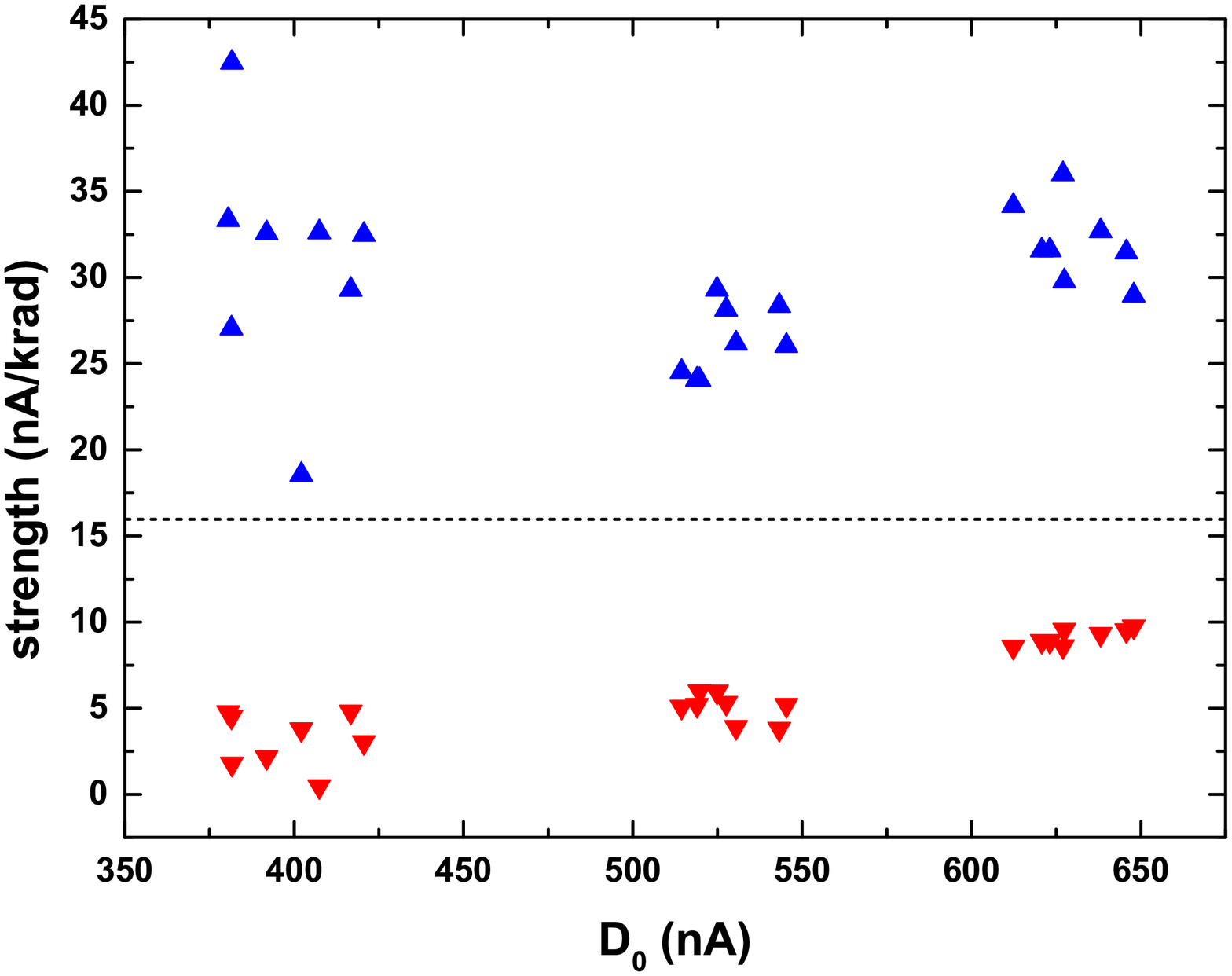}
%\caption{}
\label{}
%\end{subfigure}%
~
%\\
%\begin{subfigure}[b]{0.45\textwidth}
\centering
\includegraphics[width=0.45\textwidth]{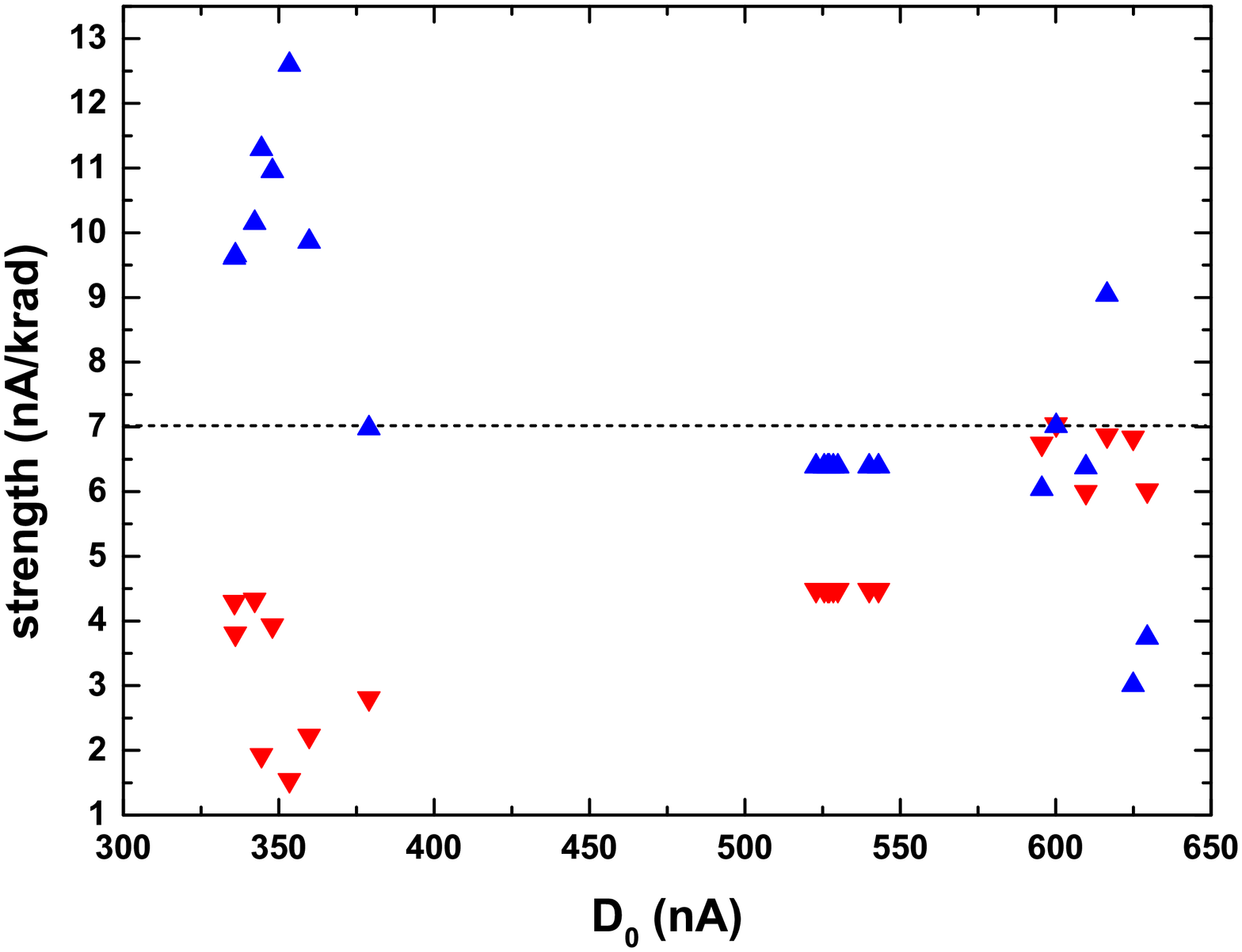}
%\caption{}
\label{}
%\end{subfigure}%
\caption{
(color online) (Initial) strength of the charge- and proton-induced synergistic effects (red and blue) as
a function of the initial DD for the (a) low and (b) high dose rate.
The strenghs for the ID are shown as the dashed lines.
}\label{fig:strength}
\end{figure*}

\subsection{Dependence of the exponential synergistic effect on the gamma ray dose rate:
proton-induced defect passivation
near the silica-silicon interface
}

Having understood the origin of the linear synergistic effect,
we now investigate what is the origin of the exponential synergistic effect,
$- b (1 - e^{- c x})$ in Eq. (4).
{As we can see in Fig. 7 (b)},
the amplitude of the exponential synergistic term ($b$)
shows an evident ELDRS effect.
{For the lower dose rate, $b$ has a higher value of about 13.5nA,
while for the higher dose rate, $b$ has a smaller value of about 3.0nA}.
Accordingly, the enhancement factor is 4.5, which is about two times of
that for the pure ID, see Figs. 4 (a) and (e).
As mentioned before, protons play a central role
in the ELDRS effect of the ID.~\cite{pease2008eldrs}
Thus, the strong ELDRS effect
of the exponential synergistic effect suggests proton
is involved in the process.
To test this assumption, we did further analysis in the following.

Some experiments and theories have shown that,
hydrogen can penetrate the a-SiO2/Si interface~\cite{Pantelides2000Reactions}
 and diffuse into Si~\cite{Sopori1996Hydrogen,Hanoka1986Hydrogen,VandeWalle1988Theory}.
Further, other experiments and theories also show that,
hydrogen is capable of passivating
various types of acceptor-like defects and extended defects in Si~\cite{Sopori1996Hydrogen,Corbett1991Hydrogen,Hanoka1986Hydrogen,
Witczak1998Space,Pearton1985Properties,Zhang2001Hydrogen,
Johnson1985Mechanism,Assali1985Microscopic,Mathiot1989Modeling,
Pearton1992Hydrogen,Sana1994Gettering,Bourret-Sicotte2017Shielded}.
Many types of complex have been proposed, including $V_2H_6$, $V_2H_8$, $VH_2$, $VH$, etc~\cite{Corbett1991Hydrogen,Zhang2001Hydrogen,gerasimenko1978infrared}.
The representative reactions can be described as %in form of the equation}
\begin{equation}
V_2 H_n^-  + H^+ \rightleftharpoons V_2 H_{n+1},
\end{equation}
where $V_2 H_{n+1}$ is electrically non-active, {see Fig. 1(b)}.
These reactions
remove the band-gap levels in silicon thus reduce the %charge carriers’
recombination rate (decrease the displacement damage).

Supposing the concentration of $V_2H_n$
contributes a base current of $D_1$. %or $b$.
From {Figs. 5 and 7(b)}, it is clear that,
$D_1$ (about several nA) is much {less} than
$D_0$ (about hundreds of nA).
This means that,  the concentration of $V_2H_n$ is much fewer than
the total amount of the neutron-induced defects in silicon.
Thus, not many protons are required to totally anneal the former defects.
The relation also means that, the reaction of Eq. (8) can cause sensible changes to the concentration of $V_2H_n$.
In other words, 
the concentration is an explicit function of time, $V_1(t)$.
From Eq. (8), 
the decay rate of $V_1(t)$ is $d [V_1](t)/dt= - k_p [H] [V_1](t)$,
where $k_p$ measures the reaction rate between the defects and protons.
Integrating the equation, the defect concentration as a function of time
({gamma ray dose}) is obtained as
\begin{equation}
V_1(t)=V_1 e^{-\beta x}~,
\end{equation}
where we have replaced $k_p [H] t$ with $\beta x$.
Accordingly, the input bias current changes from $D_1$
to $D_1 e^{-\beta x}$,
which results in a current decrease of
\begin{equation}\label{eq:biascurrent}
\Delta I_{iB}^{SE2} = - D_1 (1 - e^{- \beta x})~.
\end{equation}
Eq. (\ref{eq:biascurrent}) has the same form as the exponential term in Eq.~(\ref{eq:fitting}),
with the amplitude $D_1$ corresponding to the fitting factor $b$
and
the effective decay rate $\beta$ corresponding to the fitting factor $c$.

{The reasons for the strong ELDRS effect of the amplitude of the exponential
synergistic effect %PDI
can be explained as following.
}
{1) For the lower dose rate, more protons are generated in silica as a result of
the ELDRS effect of the ID,
so more protons can diffuse into silicon and leads to a larger passivation.
2) For the lower dose rate, there is a much longer time for protons
to diffuse into silicon, which further increases the difference of the amount of the protons in silicon.
So, it can be concluded that, the exponential synergistic effect stems from
the defect passivation near the silica-silicon interface,
which is induced by %reactions with
protons diffusing from the silica, see Fig. 1(b).}

The variable $\beta$ or $c$
shows no evident dependence on $D_0$ {or the gamma ray dose rate}. %nor $D_1$.
This is because it measures an intrinsic interaction strength,
%similar to $\alpha$,
whose value does not depend on the concentrations of the reactants 
($V_2H_n^-$ and $H^+$),
which are determined by $D_0$ or the dose rate.

\subsection{The relative strength of the two synergistic effects
}

\begin{figure*}[!t]
\centering
%\begin{subfigure}[t]{0.46\textwidth}
\centering
\includegraphics[width=0.46\textwidth]{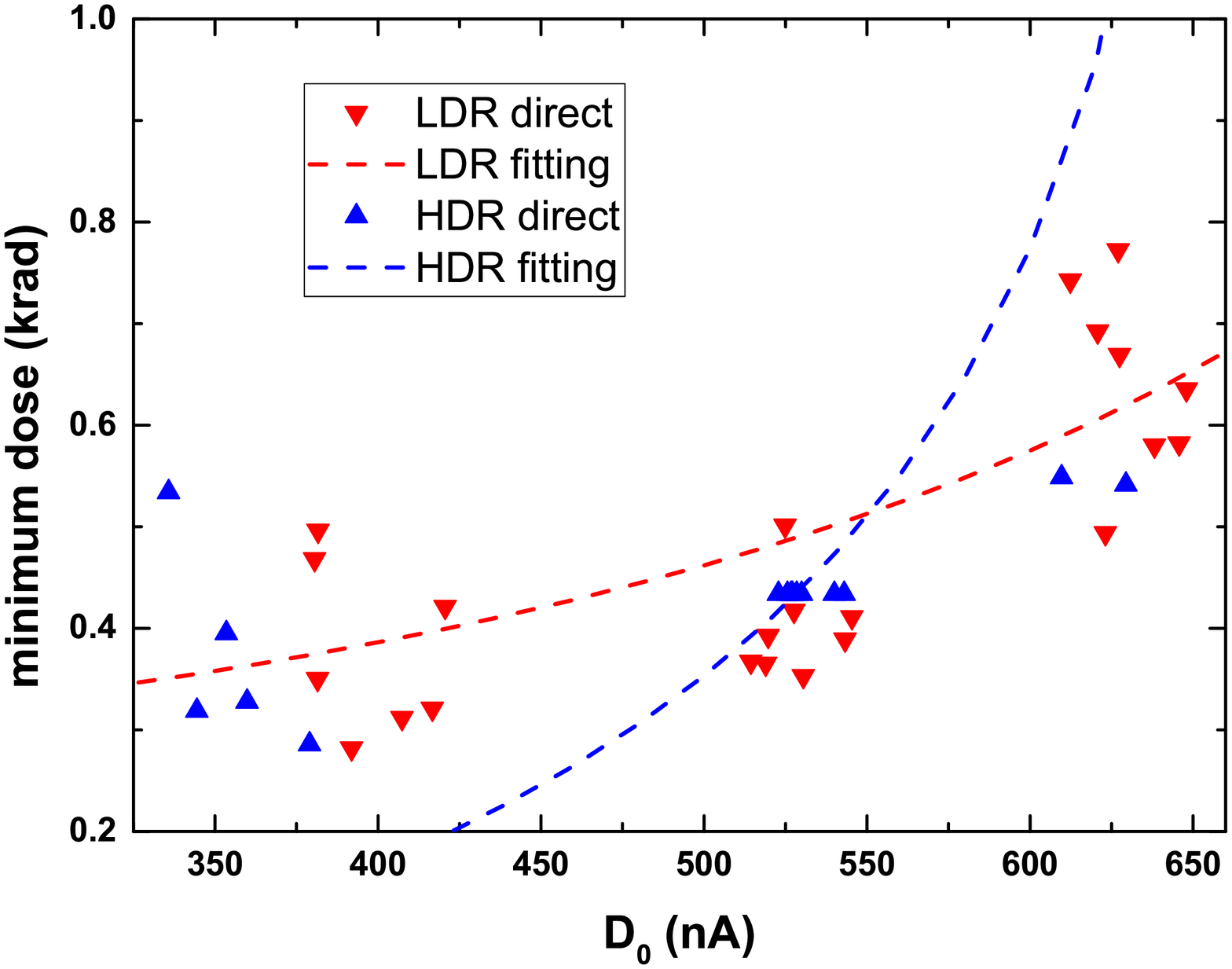}
%\caption{}
\label{}
%\end{subfigure}%
~
%\begin{subfigure}[t]{0.45\textwidth}
\centering
\includegraphics[width=0.45\textwidth]{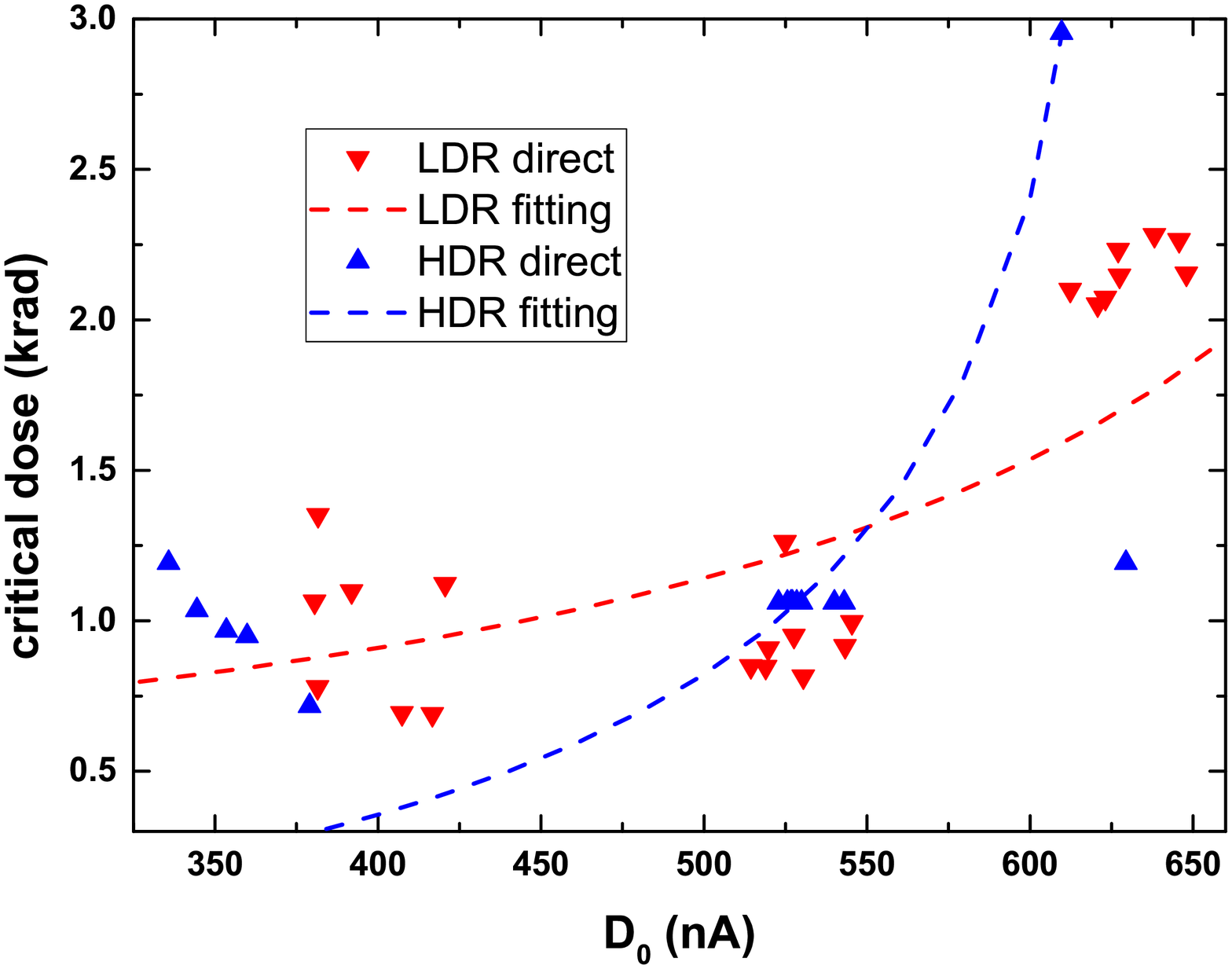}
%\caption{}
\label{}
%\end{subfigure}%
\caption{(color online)
The (a) minimal and (b) critical damage dose as a function of the initial DD.
Dots are for values directly read from the data
and curves are for results calculated by Eq. (14) or (16).
Red is for the low dose rate and blue is for the high dose rate.
}\label{fig:minimaldamage}
\end{figure*}

By combining Eqs.~(2), (\ref{eq:linearannealing}), and (\ref{eq:biascurrent}),
the practical neutron-gamma damage
is described by the following equation:
\begin{equation}\label{eq:deviationcurrent}
\Delta I_{iB}^{syn} = D_0 + k_0 x - \alpha D_0^2 x - D_1 (1 - e^{- \beta x})~.
\end{equation}
It is clear that, the practical damage deviates from the simple summed one ($D_0+k_0x$)
by two synergistic effects, namely
the carrier-induced linear annihilation, $-\alpha D_0^2x$,
and the proton-induced exponential passivation, $ - D_1 (1 - e^{- \beta x})$.
This is very different from the existing mechanism: %of synergistic effect:
Coulomb interaction induced change in the concentration of charge carriers in silicon.
Now an important question would arise: How strong are
the two synergistic effects?
Should them be weak enough %to the extent
that they can be neglected?
To answer this question, we derive the changing rates of
the input bias currents by
taking derivative of $\Delta I_{iB}^{syn}$ in Eq.~(\ref{eq:deviationcurrent}) with respect to $x$.
The result is
\begin{equation}\label{eq:derivative}
\frac{d \Delta I_{iB}^{syn}}{d x} = k_0 - \alpha D_0^2 - \beta D_1 e^{-\beta x}~.
\end{equation}
Thus, the strength for each effect can be defined as the corresponding slope,
which reads $k_0$, $\alpha D_0^2$, and $\beta D_1 e^{-\beta x}$ for the ID,
the carrier-indued linear annihilation,
and the proton-induced exponential passivation, respectively.
It is seen that, the strength for the proton-induced passivation decays with
the time. In the following, we would like to discuss the initial strength $\beta D_1$.
The values of $\alpha D_0^2$ and $\beta D_1$ are plotted
in Fig. 8 (a) and (b)
for the low and high dose rate, respectively.
The values of $k_0^L$ and $k_0^H$ are indicated by the dashed lines.
For the low dose rate case, it is seen that,
the strength of the carrier-induced annihilation, which increases with the initial DD,
is a little smaller than the strength of the ID,
while the initial strength of proton-induced passivation
is stronger than the stength of the ID.
The cases for the three typcial samples
radiated with the low dose rate can be clearly seen in Fig. 6.
For the high dose rate case, both the carrier-induced annihilation and the
proton-induced passivation are comparable with the ID.
So, although the two negative synergistic effects (of the order of nA) are
much smaller than the initial DD (of the order of hundreds of nA),
they are comparable with the ID.

It is interesting that,
the initial strength of the proton-induced passivation ($\beta D_1$) is larger than
the strengh of the carrier-induced annihilation ($\alpha D_0^2$),
although the population of the defects involved in the former ($\propto D_1$)
is about two order of magnitudes smaller than the
population of the defects involved in the latter ($\propto D_0$).
The reason is that, the reactions in the proton-induced passivation
are much easier
than the reactions in the carrier-induced annihilation,
because the $V_2H_n^-$ defects in the former contain dangling bonds and the binding energy with protons
are negative~\cite{Corbett1991Hydrogen,Zhang2001Hydrogen,gerasimenko1978infrared}.

\subsection{Conditions for the `tick'-like profiles and specific damages}

The `tick'-like damage-dose profiles in Figs. 4 and 5
are essential to obtain the two negative synergistic effects.
In this last subsection, we will try to investigate
under which neutron fluence such a profile can arise.
To further verify %the validity of
the proposed model (Eq. (11)),
we will also calculate under what gamma ray dose the damage will become the minimal or
the same as the initial one, and compare the results with
the experiment data.

From Eq. (12), it is readily seen that,
if the sum of the (initial) strengths of the two synergistic effects
are larger than the strength of the ID,
the positive ID term will be overwhelmed
and the damage-dose curves will show the declining behaviors.
For large enough gamma ray dose, the strength of the proton-induced passivation
almost vanishes and
the damage-dose curves show ascending behaviors,
provided the strength of the ID is stronger than that of the linear annealing.
%%%%
So, the conditions to display a `tick'-like profile is that,
the ID is smaller than the sum of two annealing effects
but larger than the linear annihilation effect.
In the view of initial DD (neutron fluence), this means that
the carrier-induced annealing should be stronger
than the difference between the ID and the proton-induced passivation,
but smaller than the ID itself,
\begin{equation}
\sqrt{\alpha^{-1}(k_0 - \beta D_1)} <  D_0 < \sqrt{\alpha^{-1} k_0}.
\end{equation}
{For the low dose rate, we obtain $D_0<$872.9 nA for the appearance
of the ascending behavior
at large gamma ray dose and no requirement for the declining behaviors at the small dose.
For the high dose rate, we obtain 237.2 nA $< D_0 <$ 661.4 nA
for the appearance of the `tick'-like profiles.
These conditions are consistent with the experimental data in Fig. 4.
On the other hand, we have obtained monotonously decreasing
damage-dose profiles with high gamma ray dose rate for $D_0$ of about 1000 nA
(using neutron fluence of 1$\times$10$^{14}$/cm$^2$).}

At a gamma ray dose where the derivative of the input bias current on the
gamma ray dose becomes zero, the damage reaches its minimal value.
From Eq.~(\ref{eq:derivative}), the gamma ray dose is obtained as
\begin{equation}
\beta x_{min}=\ln \left(\frac{\kappa}{k}\right)~,
\end{equation}
where $k=k_0 - \alpha D_0^2$ and $\kappa=\beta D_1$.
Physically, this equation means that, at the specific dose of $x_{min}$
the decrease rate of the base current due to the two kinds of defect annealing
in silicon equals to the increase rate of the base current
due to the defect generation on the interface.
Inserting this condition into Eq.~(\ref{eq:deviationcurrent}), 
the minimal current is obtained as
\begin{equation}
I_{iB}^{min}=D_0 + D_1 \left[\frac{k}{\kappa}\ln \left(\frac{\kappa}{k}\right)-\left(1-\frac{k}{\kappa}\right)\right]~.
\end{equation}
The calculated curve {using Eq. (14)}
is shown in Fig.~\ref{fig:minimaldamage} (a) for both
the high and low dose rates.
Also shown are the dose values for each sample,
which is directly read from the data in Fig. 4.
For the low dose rate case,
the calculated curve is found to have good agreements with the experimental data,
further implying the validity of the proposed model.
However, for the high dose rate case, the agreement is only good for the curves with initial DD of about 530nA in Fig. 4 (g).
This is because the dose step of 0.5krad (total dose of 5krad) is too large (small) for the curves with $D_0$ of about 350nA (610nA) in Fig. 4f (4h) and the true dose values can be easily missed.
From the model predictions, it is also seen that, for both the high and low dose rates, the dose condition
for the minimal damage increases for the increasing $D_0$.

For certain larger gamma ray dose, the damage returns to
the initial value before gamma radiation (i.e.,
$D_0$), which we can call as a critical dose.
Solving Eq.~(\ref{eq:deviationcurrent}) with the condition of $\Delta I_{iB}^{syn}=D_0$, the critical gamma ray dose is obtained as
\begin{equation}\label{eq:equaldamage}
\beta x_c = \frac{\kappa}{k}+ W\left[-\frac{\kappa}{k} e^{-\frac{\kappa}{k}}\right]~,
\end{equation}
where $W$ is Lambert-W function or product logarithm.
Physically, this equation means that, at the specific dose 
of $x_c$ the base current due to the two kinds of defect annealing
in silicon equals to the base current
due to the defect generation on the interface.
Again the calculated curve using Eq. (16) and the experimental data
are shown in Fig.~\ref{fig:minimaldamage} (b),
which are found to be in good agreement with each other
for the low dose rate case. 
For the high dose rate case,
the agreement is bad for the samples with $D_0$ of about 350nA (610nA),
due to the too large dose step (too small total dose).
It can be seen that, for both the high and low dose rates,
the critical dose depends monotonously on the initial DD.

It has been noticed that, the ratio of the strength of the proton-induced passivation ($\kappa = \beta D_1$)
and the sum of the strengths of the ID and the charge-induced annihilation ($k = k_0 - \alpha D_0^2$) plays a crucial role in all the critical parameters
in Eqs. (14)-(16).
This result
reveals the significance of the cooperation of the
ID and the two synergistic effects.

\section{Conclusion}

In summary, we have systematically studied {the behavior and mechanism of}
the synergistic effects of neutron and gamma ray radiation by performing successive neutron-gamma radiation experiments
on input-stage PNP transistors in operational amplifier LM324N. %with various initial DD.
We find that the measured input bias current obey a `tick'-like damage-dose curve.
Two negative synergistic effects, both of which are comparable with the ID itself,
have been derived from the experimental observation.
The first one is caused by a carrier-induced defect annihilation in silicon;
it displays as a linear function of the gamma ray dose whose slope
depends quadratically on the initial DD of the samples.
The second one is caused by a proton-induced defect passivation near the silica/silicon
interface; it displays as an exponential function of the gamma ray dose whose
amplitude shows a very strong ELDRS effect.
The validity of the proposed model is also verified by the prediction of
the dose for the minimal and critical synergistic damage.
Our proposed mechanism demonstrates that, the ID can influence DD
by decreasing the concentration of defects in it, therefore, this technique
can be applied to repair devices used in
the space and other extreme environments.

The authors thank Professor Chun Zheng of Institute of Nuclear Physics and Chemistry, CAEP for his kindly help in neutron radiation experiments.
This work was supported by the Science Challenge Project under Grant No. TZ2016003-1 and NSFC under Grant Nos. 51672023; 11634003; U1530401.

\bibliography{Refs-interaction-new}

%\begin{tocentry}
%
%\includegraphics[width=\linewidth]{TOC-0308}
%
%\end{tocentry}

\end{document}